\documentclass[journal,twocolumn,10pt]{IEEEtran}
\usepackage[]{times}
\usepackage{epsfig}
\usepackage{subfigure}
\usepackage{amsmath}
\usepackage{amssymb}
\usepackage{graphicx}
\usepackage{subfigure}
\usepackage{multirow}
\usepackage{flushend}
\usepackage{url}
\usepackage{color}
\newcommand{\bega}{\begin{eqnarray}}
\newcommand{\ega}{\end{eqnarray}}
\newcommand{\bb}{\begin{equation}}
\newcommand{\ee}{\end{equation}}
\newcommand{\supp}{\mbox{supp}}
\newtheorem{defn} {Definition}
\newtheorem{te}{Theorem}
\newtheorem{lema}{Lemma}
\newtheorem{cor}{Corollary}
\newtheorem{ex}{Example}

\begin{document}

\title{Provably efficient instanton search algorithm for LP decoding of LDPC codes over the BSC}

\author{Shashi~Kiran~Chilappagari,~\IEEEmembership{Student~Member,~IEEE,}
\thanks{Manuscript received \today. S. K. Chilappagari [shashic@ece.arizona.edu] is with the ECE Department, University of Arizona, Tucson, AZ, 85721, USA. }
~Michael Chertkov,~\IEEEmembership{Member,~IEEE}
\thanks{M. Chertkov [chertkov@lanl.gov] is with Theory Division \& CNLS, LANL, Los Alamos, NM, 87545 USA.}
and ~Bane Vasic, ~\IEEEmembership{Senior Member,~IEEE}
\thanks{B. Vasic [vasic@ece.arizona.edu] is with the ECE Department, University of Arizona, Tucson, AZ, 85721, USA.}}

\markboth{Submitted to IEEE Transactions on Information Theory, September 2008}
{Chilappagari, Chertkov \& Vasic: Provably Efficient Instanton Search Algorithm}
\maketitle

\begin{abstract}
We consider Linear Programming (LP) decoding of a fixed Low-Density Parity-Check (LDPC) code over the Binary Symmetric Channel (BSC). The LP decoder fails when it outputs a pseudo-codeword which is not a codeword. We design an efficient algorithm termed the Instanton Search Algorithm (ISA) which, given a random input, generates a set of flips called the BSC-instanton. We prove that: (a) the LP decoder fails for any set of flips with support vector including an instanton; (b) for any input, the algorithm outputs an instanton in the number of steps upper-bounded by twice the number of flips in the input. Repeated sufficient number of times, the ISA outcomes the number of unique instantons of different sizes. 
\end{abstract}



\section{Introduction}

The significance of Low-Density Parity-Check (LDPC) codes \cite{61Gal} is in their capacity-approaching performance when decoded using low complexity iterative algorithms, such as Belief Propagation (BP) \cite{61Gal,richardsonurbanke}. Properly chosen sequence of LDPC codes  can be made asymptotically good, i.e. iterative decoding guarantees exponential decay of error probability in the code length $n$ when the noise is below a finite threshold. Iterative decoders operate by passing messages along the edges of a graphical representation of a code known as the Tanner graph \cite{tanner81}, and are optimal when the underlying graph is a tree. However, the decoding becomes sub-optimal in the presence of cycles, and hence the above threshold statement is of a limited practical use for the analysis of a fixed code. The linear programming (LP) decoding introduced by Feldman \textit{et al.} \cite{05FWK}, is another sub-optimal algorithm for decoding LDPC codes, which has higher complexity but is more amenable to analysis. 

The typical performance measures of a decoder (either LP or BP) for a fixed code are the Bit-Error-Rate (BER) or/and the Frame-Error-Rate (FER) as functions of the Signal-to-Noise Ratio (SNR). A typical BER/FER vs SNR curve consists of two distinct regions. At small SNR, the error probability decreases rapidly with the SNR, and the curve forms the so-called \textit{water-fall} region. The decrease slows down at moderate values turning into the \textit{error-floor} asymptotic at very large SNR \cite{03Rich}. This transient behavior and the error-floor asymptotic originate from the sub-optimality of the decoding, i.e., the ideal maximum-likelihood (ML) curve would not show such a dramatic change in the BER/FER with the SNR increase.

After the formulation of the problem by Richardson \cite{03Rich}, a significant effort has been devoted to the analysis of the error floor phenomenon.  Given that the decoding sub-optimality is expressed in the domain where the error probability is small, the troublesome noise configurations leading to decoding failures and controlling the error-floor asymptotic are extremely rare, and  analytical rather than simulation methods for their characterization are necessary. It is worth noting here that most of the analytical methods developed in the theory of iterative decoding have focused on ensembles of codes rather than a given fixed code.

The failures of iterative decoding over the binary erasure channel (BEC) are well understood in terms of combinatorial objects known as stopping sets \cite{di}. For iterative decoding on the Additive White Gaussian Noise (AWGN) channel and the BSC, the decoding failures have been characterized in terms of trapping sets \cite{03Rich,06CSV} and pseudo-codewords \cite{wiberg,03KV,05VK}. Richardson \cite{03Rich} introduced the notion of trapping sets and proposed a semi-analytical method to estimate the FER performance of a given code on the AWGN channel in the error floor region. The method was successfully applied to hard decision decoding over the BSC in \cite{06CSV}. The approach of \cite{03Rich} was further refined by Stepanov \textit{et al.} \cite{PRLswithBane}, using \textit{instantons}.  Pseudo-codewords were first discussed in the context of iterative decoders using computation trees \cite{wiberg} and later using graph covers \cite{03KV,05VK}. Pseudo-codeword distributions were found for the special cases of codes from Euclidean and projective planes \cite{07SV}. A detailed analysis of the pseudo-codewords was presented by Kelley and Sridhara \cite{kellysridhara}, who discussed the bounds on pseudo-codeword size in terms of the girth and the minimum left-degree of the underlying Tanner graph. The bounds were further investigated by Xia and Fu \cite{08XF}. Pseudo-codeword analysis has also been extended to the convolutional LDPC codes by Smarandache \textit{et al.} \cite{06SPVC}. (See also \cite{Pascalswebsite} for an exhaustive list of references for this and related subjects.)

Pseudo-codewords can be also used to understand the failures of the LP decoder \cite{05FWK}. The pseudo-codewords for the LP decoder are equivalent to stopping sets for the case of the BEC. For the AWGN channel, the pseudo-codewords of the LP decoder are related to the pseudo-codewords arising from graph covers \cite{05VK}. In fact, in \cite{05VK} Vontobel and Koetter have also pointed out relations between pseudo-codewords arising from graph covers and trapping sets.

Closely related to the pseudo-codewords and the trapping sets are the noise configurations that lead to decoding failures which are termed as instantons \cite{PRLswithBane}. Finding the instantons is a difficult task which so far admitted only heuristic solutions \cite{06CSV,07C}. In this regard, the most successful (in efficiency) approach, coined the Pseudo-Codeword-Search (PCS) algorithm, was suggested for the LP decoding performing over the continuous channel in \cite{08CS} (with Additive White Gaussian Noise (AWGN) channel used as an enabling example). Given a sufficiently strong random input, the outcome of the PCS algorithm is an instanton. The resulting distribution of the instantons (or respective pseudo-codewords) thus provides a compact and algorithmically feasible characterization of the AWGN-LP performance of the  given code.

In this paper, we consider pseudo-codewords and instantons of the LP decoder for the BSC. We define the \textit{BSC-instanton} as a noise configuration which the LP decoder decodes into a pseudo-codeword distinct from the all-zero-codeword while any reduction of the (number of flips in) BSC-instanton leads to the all-zero-codeword. Being a close relative of the BP decoder (see \cite{03WJ,06CC} for discussions of different aspects of this relation), the LP decoder appeals due to the following benefits: (a) it has ML certificate i.e., if the output of the decoder is a codeword, then the ML decoder is also guaranteed to decode into the same codeword; (b) the output of the LP decoder is discrete even if the channel noise is continuous (meaning that problems with numerical accuracy do not arise); (c) its analysis is simpler due to the readily available set of powerful analytical tools from the optimization theory; and (d) it allows systematic sequential improvement, which results in decoder flexibility and feasibility of an LP-based ML for moderately large codes \cite{Yedidia,06Fel}. While slower decoding speed is usually cited as a disadvantage of the LP decoder, this potential problem can be significantly reduced, thanks to the recent progress in smart sequential use of LP constraints \cite{06TS} and/or appropriate graphical transformations \cite{06Fel,06DW,07CS}.

The two main contributions of this paper are: (1) characterization of all the failures of the LP decoder over the BSC in terms of the instantons, and (2) a provably efficient Instanton Search Algorithm (ISA). Following the idea by Chertkov and Stepanov \cite{08CS}, for a given a random binary $n$-tuple, the ISA generates a BSC-instanton, that is guaranteed to be decoded by the LP decoder into a pseudo-codeword distinct from the all-zero-codeword. Our ISA constitutes a significantly stronger algorithm than the one of \cite{08CS} due to its property that it outputs an instanton in the number of steps upper-bounded by twice the number of flips in the original configuration the algorithm is initiated with.

The rest of the paper is organized as follows. In Section \ref{preliminaries}, we give a brief introduction to the LDPC codes, LP decoding and pseudo-codewords. In Section \ref{medianinstanton}, we introduce the BSC-specific notions of the pseudo-codeword weight, medians and instantons 
(defined as special set of flips), their costs, and we also prove  some set of useful lemmata emphasizing  the significance of the instanton analysis. In Section \ref{algorithm}, we describe the ISA and prove our main result concerning bounds on the number of iterations required to output an instanton.
We present the ISA test, as applied to the $[155,64,20]$ Tanner code \cite{tannercode}, in Section \ref{results}. We summarize our results and conclude by listing some open problems in Section \ref{conclusion}.

\section{Preliminaries: LDPC Codes, LP Decoder and Pseudo-Codewords}\label{preliminaries}

In this Section, we discuss the LP decoder and the notion of pseudo-codewords.
We adopt the formulation of the LP decoder and the terminology from
\cite{05FWK}, and thus the interested reader is advised to refer to \cite{05FWK}
for more details.

Let $\mathcal{C}$ be a binary LDPC code defined by a Tanner graph $G$ with two sets of nodes: the set of variable nodes $V=\{1,2,\ldots,n\}$ and the set of check nodes $C=\{1,2,\ldots,m\}$. The adjacency matrix of $G$ is $H$, a parity-check matrix of $\mathcal{C}$, with $m$ rows corresponding to the check nodes and $n$ columns corresponding to the variable nodes. A binary vector $\mathbf{c}=(c_1,\ldots,c_n)$ is a codeword iff $\mathbf{c}H^{T}=\mathbf{0}$. The support of a vector $\mathbf{r}=(r_1,r_2,\ldots,r_n)$, denoted by $\supp(\mathbf{r})$, is defined as the set of all positions $i$ such that $r_i\neq 0$.

We assume that a codeword $\mathbf{y}$ is transmitted over a discrete symmetric memoryless channel and is received as $\mathbf{\hat{y}}$. The channel is
characterized by $\Pr[\hat{y}_i|y_i]$ which denotes the probability that $y_i$ is received as $\hat{y}_i$. The negative log-likelihood ratio (LLR) corresponding to the variable node $i$ is given by
\[
\gamma_i=\log\left(\frac{\Pr(\hat{y}_i| y_i=0)}{\Pr(\hat{y}_i| y_i=1)}\right).
\]
The ML decoding of the code $\mathcal{C}$ allows a convenient LP formulation in
terms of the \textit{codeword polytope} $\mbox{poly}(\mathcal{C})$ whose
vertices correspond to the codewords in $\mathcal{C}$. The ML-LP decoder finds
$\mathbf{f}=(f_1,\ldots,f_n)$ minimizing the cost function
$\sum_{i=1}^{n}\gamma_if_i$ subject to the $\mathbf{f}\in
\mbox{poly}(\mathcal{C})$ constraint. The formulation is compact but
impractical because of the number of constraints exponential in the code
length.

Hence a \textit{relaxed} polytope is defined as the intersection of all the
polytopes associated with the local codes introduced for all the checks of the
original code. Associating $(f_1,\ldots,f_n)$ with bits of the code we require
\begin{equation}\label{eq1}
0 \leq f_i \leq 1, ~~\forall i \in V
\end{equation}
For every check node $j$, let $N(j)$ denote the set of variable nodes which are
neighbors of $j$. Let $E_j=\{T \subseteq N(j): |T| \mbox{~is even}\}$. The
polytope $Q_j$ associated with the check node $j$ is defined as the set of
points $(\mathbf{f},\mathbf{w})$ for which the following constraints hold
\begin{eqnarray}
&0 \leq w_{j,T} \leq 1,& \forall T \in E_j \\
&\sum_{T \in E_j} w_{j,T}=1& \\
&f_i=\sum_{T \in E_j, T \ni i }w_{j,T},& \forall i \in N(j) \label{eq4}
\end{eqnarray}
Now, let $Q=\cap_j Q_j$ be the set of points $(\mathbf{f},\mathbf{w})$ such that
(\ref{eq1})-(\ref{eq4}) hold for all $j \in C$. (Note that $Q$, which is also referred to as the fundamental polytope \cite{03KV,05VK}, is a function of the Tanner graph $G$ and consequently the parity-check matrix $H$ representing the code $\mathcal{C}$.) The Linear Code
Linear Program (LCLP) can be stated as
\[
\min\limits_{(\mathbf{f},\mathbf{w})} \sum_{i \in V}\gamma_i f_i, \mbox{~s.t.~} (\mathbf{f},\mathbf{w}) \in Q.
\]
For the sake of brevity, the decoder based on the LCLP is referred to in the
following as the LP decoder. A solution $(\mathbf{f},\mathbf{w})$ to the
LCLP such that all $f_i$s and $w_{j,T}$s are integers is known as an integer
solution. The integer solution represents a codeword \cite{05FWK}. It was also
shown in \cite{05FWK} that the LP decoder has the ML certificate, i.e., if the
output of the decoder is a codeword, then the ML decoder would decode into the
same codeword. The LCLP can fail, generating an output which is not a codeword.

The performance of the LP decoder can be analyzed in terms of the
pseudo-codewords, originally defined as follows:
\begin{defn}\cite{05FWK}
\textit{Integer pseudo-codeword} is a vector $\mathbf{p}=(p_1,\ldots,p_n)$ of
non-negative integers such that, for every parity check $j \in C$,
the neighborhood $\{p_i: i \in N(j)\}$ is a sum of local codewords.
\end{defn}
Alternatively, one may choose to define a \textit{re-scaled pseudo-codeword},
$\mathbf{p}=(p_1,\ldots,p_n)$ where $0 \leq p_i \leq 1, \forall i
\in V$, simply equal to the output of the LCLP. In the following, we
adopt the re-scaled definition. 

A given code $\mathcal{C}$ can have different Tanner graph representations and consequently potentially different fundamental polytopes. Hence, we refer to the pseudo-codewords as corresponding to a particular Tanner graph $G$ of $\mathcal{C}$.

It is also appropriate to mention here that the LCLP can be viewed as the zero
temperature version of  BP-decoder looking for the global minimum of the
so-called Bethe free energy functional \cite{03WJ}.

\section{Cost and Weight of Pseudo-codewords, Medians and Instantons}\label{medianinstanton}

Since the focus of the paper is on the pseudo-codewords for the BSC, in this Section we introduce some terms, e.g. instantons and medians, specific to the BSC. We will also prove here some preliminary lemmata which will enable subsequent discussion of the ISA in the next Section.

The polytope $Q$ is symmetric and looks exactly the same from all codewords (see e.g. \cite{05FWK}). Hence we assume that the all-zero-codeword is
transmitted. The process of changing a bit from $0$ to $1$ and vice-versa is known as flipping. The BSC flips every transmitted bit with a certain probability. We therefore call a noise vector with support of size $k$ as having $k$ flips.

In the case of the BSC, the likelihoods are scaled as
\[
\gamma_i=\left\{ \begin{array}{cl}
1,& \mbox{~if~} y_i=0; \\
-1,&\mbox{~if~} y_i=1. \end{array}\right.
\]

Two important characteristics of a pseudo-codeword are its cost and weight.
While the cost associated with decoding to a pseudo-codeword has already been
defined in general, we formalize it for the case of the BSC as follows:
\begin{defn}\label{def2}
The cost associated with LP decoding of a binary vector $\mathbf{r}$ to a
pseudo-codeword $\mathbf{p}$ is given by
\begin{equation}
C(\mathbf{r},\mathbf{p})=\sum_{i\notin \supp(\mathbf{r})}p_i - \sum_{i \in \supp(\mathbf{r})}p_i.
\end{equation}
\end{defn}
If $\mathbf{r}$ is the input, then the LP decoder converges to the pseudo-codeword $\mathbf{p}$ which has the least value of $C(\mathbf{r},\mathbf{p})$. The cost of decoding to the all-zero-codeword is zero. Hence, a binary vector $\mathbf{r}$ does not converge to the all-zero-codeword if there exists a pseudo-codeword $\mathbf{p}$ with $C(\mathbf{r},\mathbf{p}) \leq 0$.

\begin{defn}\cite[Definition 2.10]{kellysridhara}
Let $\mathbf{p}=(p_1,\ldots,p_n)$ be a pseudo-codeword distinct from the
all-zero-codeword. Let $e$ be the smallest number such that the sum of the $e$
largest $p_i$s is at least $\left(\sum_{i \in V}p_i\right)/2$. Then, the BSC
\textit{pseudo-codeword weight} of $\mathbf{p}$ is
\[
w_{BSC}(\mathbf{p})=\left\{ \begin{array}{cl}
2e,& \mbox{~if~} \sum_{e}p_i=\left(\sum_{i \in V}p_i\right)/2; \\
2e-1,&\mbox{~if~} \sum_{e}p_i>\left(\sum_{i \in V}p_i\right)/2. \end{array}\right.
\]
\end{defn}

The minimum pseudo-codeword weight of $G$ denoted by $w_{min}^{BSC}$ is the minimum over all the non-zero pseudo-codewords of $G$. The parameter $e=\lceil \left(w_{BSC}(\mathbf{p})+1\right)/2 \rceil$ can be interpreted as the least number of bits to be flipped in the all-zero-codeword such that the resulting vector decodes to the pseudo-codeword $\mathbf{p}$. (See e.g. \cite{01Forney} for a number of illustrative examples.)

\textit{Remark:} Feldman \textit{et al.} in \cite{05FWK} defined \textit{weight} of a pseudo-codeword, the \textit{fractional distance} and the \textit{max-fractional distance} of a code in terms of the projected polytope $\overline{Q}$ (the interested reader is referred to \cite{05FWK} for explicit description of $\overline{Q}$). To differentiate the two definitions, we term the ``weight'' defined by Feldman \textit{et al.} as \textit{fractional weight} and denote it by $w_{frac}$. For a point $\mathbf{f}$ in $\overline{Q}$, the fractional weight of $\mathbf{f}$ is defined as the L1-norm, $w_{frac}(\mathbf{f})= \sum_{i \in V} f_i$ and the max-fractional weight of $\mathbf{f}$ is defined as the fractional weight normalized by the maximum $f_i$ value i.e., 
\[
w_{max-frac}(\mathbf{f})=\frac{w_{frac}(\mathbf{f})}{\max_i f_i}.
\] Also, if $\mathcal{V}_{\overline{Q}}$ denotes the set of non-zero vertices of $\overline{Q}$ the fractional distance $d_{frac}$ of the code is defined as the minimum weight over all vertices in $\mathcal{V}_{\overline{Q}}$. The max-fractional distance $d_{frac}^{max}$ of the code is given by
\[
d_{frac}^{max}=\min\limits_{(\mathbf{f},\mathbf{w}) \in \mathcal{V}_{\overline{Q}},\\ \mathbf{f} \neq \mathbf{0}} \left(\frac{\sum_{i \in V}f_i} {\max_i f_i}\right)
\]
It was shown in \cite [Theorem 9]{05FWK} that the LP decoder is successful if at most $\lceil d_{frac}/2 \rceil -1$ bits are flipped by the BSC, thus making $d_{frac}$ a potentially useful characteristic. Moreover, an efficient LP-based algorithm to calculate $d_{frac}$ was suggested in \cite{05FWK}. However, the error pattern with the least number of flips which the LP decoder fails to correct does not necessarily converge to the pseudo-codeword with fractional weight $d_{frac}$. Hence, we adopted the  definition of the pseudo-codeword weight from \cite{01Forney,kellysridhara}, however noticing  that it was discussed there in a different but related context of the computation tree and graph covers.  The advantage of our approach will become evident in the subsequent Sections.

The following Lemma gives a relation between $w_{BSC}^{min}$ and $d_{frac}$. 
\begin{lema}\label{lemma0}
$w_{BSC}^{min} \geq 2\lceil d_{frac}/2 \rceil -1$.
\end{lema}
\begin{proof}
The LP decoder is successful if at most $\lceil d_{frac}/2 \rceil -1$ bits are flipped by the BSC. So, the minimum number of flips in the all-zero-codeword which can cause the LP decoder to fail is $\lceil d_{frac}/2 \rceil$. If $e$ is the minimum number of flips associated with the minimum weight pseudo-codeword, then
\[
e \geq \lceil d_{frac}/2 \rceil
\]
Since, $w_{BSC}^{min} \geq 2e-1$, we have $w_{BSC}^{min} \geq 2\lceil d_{frac}/2 \rceil -1$
\end{proof}

The above lemma can be generalized to any pseudo-codeword $\mathbf{p}$ as $w_{BSC}(\mathbf{p})\geq 2\lceil w_{frac}(\mathbf{p})/2 \rceil -1$. We would like to point out that the Kelley and Sridhara in \cite{kellysridhara} have derived a similar relation between $w_{BSC}(\mathbf{p})$ and $w_{max-frac}(\mathbf{p})$ and that Sridhara in \cite{personalcomm} observed that $w_{BSC}(\mathbf{p})+1 \geq w_{max-frac}(\mathbf{p})$.

The interpretation of BSC pseudo-codeword weight motivates the following
definition of the \textit{median noise vector} corresponding to a
pseudo-codeword:
\begin{defn}
The median noise vector (or simply the median) $M(\mathbf{p})$ of a
pseudo-codeword $\mathbf{p}$ distinct from the all-zero-codeword is a binary
vector with support $S=\{i_1,i_2,\ldots,i_e\}$, such that
$p_{i_1},\ldots,p_{i_e}$ are the $e(=\lceil
\left(w_{BSC}(\mathbf{p})+1\right)/2\rceil)$ largest components of
$\mathbf{p}$.
\end{defn}
One observers that, $C\left(M(\mathbf{p}),\mathbf{p}\right) \leq 0$. From the
definition of $w_{BSC}(\mathbf{p})$, it follows that at least one median exists
for every $\mathbf{p}$. Also, all medians of $\mathbf{p}$ have  $\lceil
\left(w_{BSC}(\mathbf{p})+1\right)/2 \rceil$ flips. The proofs of the following
two lemmata are now apparent.
\begin{lema}\label{lemma1}
The LP decoder decodes a binary vector with $k$ flips into a pseudo-codeword
$\mathbf{p}$ distinct from the all-zero-codeword \textit{iff} $w_{BSC}(\mathbf{p})
\leq 2k$.
\end{lema}
\begin{lema}\label{lemma2}
Let $\mathbf{p}$ be a pseudo-codeword with median $M(\mathbf{p})$ whose support
has cardinality $k$. Then $w_{BSC}(\mathbf{p}) \in \{2k-1, 2k\}$.
\end{lema}
\begin{lema}\label{lemma3}
Let $M(\mathbf{p})$ be a median of $\mathbf{p}$ with support $S$. Then the
result of LP decoding of any binary vector with support $S' \subset S$ and
$|S'|<|S|$ is distinct from $\mathbf{p}$.
\end{lema}
\begin{proof}
Let $|S|=k$. Then by Lemma \ref{lemma2}, $w_{BSC}(\mathbf{p})\in\{2k-1,2k\}$.
Now, if $\mathbf{r}$ is any binary vector with support $S' \subset S$, then
$\mathbf{r}$ has at most $k-1$ flips and therefore by Lemma \ref{lemma1},
$w_{BSC}(\mathbf{p})\leq 2(k-1)$, which is a contradiction.
\end{proof}
\begin{lema}\label{lemma4}
If $M(\mathbf{p})$ converges to a pseudo-codeword $\mathbf{p}_{M} \neq
\mathbf{p}$, then $w_{BSC}(\mathbf{p}_{M}) \leq w_{BSC}(\mathbf{p})$. Also,
$C(M(\mathbf{p}),\mathbf{p}_{M}) \leq C(M(\mathbf{p}),\mathbf{p})$.
\end{lema}
\begin{proof}
According to the definition of the LP decoder, $C(M(\mathbf{p}),\mathbf{p}_{M})
\leq C(M(\mathbf{p}),\mathbf{p})$.

If $w_{BSC}(\mathbf{p})=2k$, then $M(\mathbf{p})$ has $k $ flips and by Lemma
\ref{lemma1}, $w_{BSC}(\mathbf{p}_{M}) \leq 2k = w_{BSC}(\mathbf{p})$.

If  $w_{BSC}(\mathbf{p})=2k-1$, then $M(\mathbf{p})$ has $k $ flips and
$C(M(\mathbf{p}),\mathbf{p})<0$. Hence, $w_{BSC}(\mathbf{p}_{M}) \leq 2k$ by
Lemma \ref{lemma1}. However, if $w_{BSC}(\mathbf{p}_{M}) = 2k$, then
$C(M(\mathbf{p}),\mathbf{p}_{M})=0$, which is a contradiction. Hence,
$w_{BSC}(\mathbf{p}_{M}) \leq 2k-1 = w_{BSC}(\mathbf{p})$.
\end{proof}

\begin{defn}
The BSC \textit{instanton} $\mathbf{i}$ is a binary vector with the following
properties: (1) There exists a pseudo-codeword $\mathbf{p}$ such that
$C(\mathbf{i},\mathbf{p})\leq C(\mathbf{i},\mathbf{0})=0$; (2) For any binary
vector $\mathbf{r}$ such that $\supp(\mathbf{r}) \subset \supp(\mathbf{i})$,
there exists no pseudo-codeword with $C(\mathbf{r},\mathbf{p})\leq 0$. The size of an instanton is the cardinality of its support.
\end{defn}
In other words, the LP decoder decodes $\mathbf{i}$ to a pseudo-codeword other than the all-zero-codeword or one finds a pseudo-codeword $\mathbf{p}$ such that $C(\mathbf{i},\mathbf{p})=0$ (interpreted as the LP decoding failure), whereas any binary vector with flips from a subset of the flips in $\mathbf{i}$ is decoded to the all-zero-codeword. It can be easily verified that if $\mathbf{c}$ is the transmitted codeword and $\mathbf{r}$ is the received vector such that $\supp(\mathbf{c}+\mathbf{r})=\supp(\mathbf{i})$, where the addition is modulo two, then there exists a pseudo-codeword $\mathbf{p'}$ such that $C(\mathbf{r},\mathbf{p'}) \leq C(\mathbf{r},\mathbf{c})$.

The following lemma follows from the definition of the cost of decoding
(the pseudo-codeword cost):
\begin{lema}\label{lemma5}
Let $\mathbf{i}$ be an instanton. Then for any binary vector $\mathbf{r}$ such
that $\supp(\mathbf{i}) \subset \supp(\mathbf{r})$, there exists a
pseudo-codeword $\mathbf{p}$ satisfying $C(\mathbf{r},\mathbf{p}) \leq 0$.
\end{lema}
\begin{proof} Since $\mathbf{i}$ is an instanton,  there exists a pseudo-codeword $\mathbf{p}$ such that
$C(\mathbf{i},\mathbf{p})\leq 0$. From Definition \ref{def2} we have,
\[
\sum_{i \notin \supp(\mathbf{i})} p_i - \sum_{i \in \supp(\mathbf{i})} p_i \leq 0. \]
Since, $\supp(\mathbf{i}) \subset \supp(\mathbf{r})$ and $p_i \geq 0, \forall i$, we have
\[
\sum_{i \notin \supp(\mathbf{r})} p_i - \sum_{i \in \supp(\mathbf{r})} p_i \leq 0,
\]
thus yielding
\[
C(\mathbf{r},\mathbf{p}) \leq 0.
\]\end{proof}
The above lemma implies that the LP decoder fails to decode every vector
$\mathbf{r}$ whose support is a superset of an instanton to the all-zero-
codeword. We now have the following corollary:

\begin{cor}
Let $\mathbf{r}$ be a binary vector with support $S$. Let $\mathbf{p}$ be a
pseudo-codeword such that $C(\mathbf{r},\mathbf{p}) \leq 0$. If all binary
vectors with support $S' \subset S$ such that $|S'| = |S|-1$, converge to
$\mathbf{0}$, then $\mathbf{r}$ is an instanton.
\end{cor}

The above lemmata lead us to the following lemma which characterizes all the
failures of the LP decoder over the BSC:
\begin{lema}\label{lemma6}
A binary vector $\mathbf{r}$ converges to a pseudo-codeword different from the
all-zero-codeword {\it iff} the support of $\mathbf{r}$ contains the support of
an instanton as a subset.
\end{lema}

The most general form of the above lemma can be stated as following: if
$\mathbf{c}$ is the transmitted codeword and $\mathbf{r}$ is the received
vector, then $\mathbf{r}$ converges to a pseudo-codeword different from
$\mathbf{c}$ \textit{iff} the $\supp(\mathbf{r}+\mathbf{c})$, where the
addition is modulo two, contains the support of an instanton as a
subset.

From the above discussion, we see that the BSC instantons are analogous to the
minimal stopping sets for the case of iterative/LP decoding over the BEC. In
fact, Lemma \ref{lemma6} characterizes all the decoding failures of the LP
decoder over the BSC in terms of the instantons and can be used to derive
analytical estimates of the code performance given the weight distribution of
the instantons. In this sense, the instantons are more fundamental than the
minimal pseudo-codewords \cite{07SV,kellysridhara} for the BSC (note, that this statement does not
hold in the case of the AWGN channel). Two minimal pseudo-codewords of the same weight can give rise to different number of
instantons. This issue was first pointed out by Forney \textit{et al.} in \cite{01Forney}. (See Examples
1, 2, 3 for the BSC case in \cite{01Forney}.) It is also worth noting that an instanton converges to a minimal pseudo-codeword. 

It should be noted that finding pseudo-codewords with fractional weight
$d_{frac}$ is not equivalent to finding minimum weight pseudo-codewords. The
pseudo-codewords with fractional weight $d_{frac}$ can be used to derive some
instantons, but not necessarily the ones with the least number of flips. However,
as $d_{frac}$ provides a lower bound on the minimum pseudo-codeword weight, it
can be used as a test if the ISA actually finds an instanton with the least
number of flips. In other words, if the number of flips in the lowest weight
instanton found by the ISA is equal to $\lceil d_{frac}/2 \rceil$, then the ISA
has indeed found the smallest size instanton.

\begin{figure*}
\centering
\subfigure[] 
{
    \label{case1}

\includegraphics[width=1.7in]{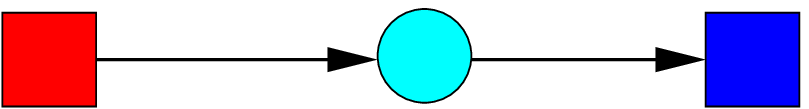}
}
\hspace{0.3in}
\subfigure[] 
{
    \label{case2}

\includegraphics[width=1.7in]{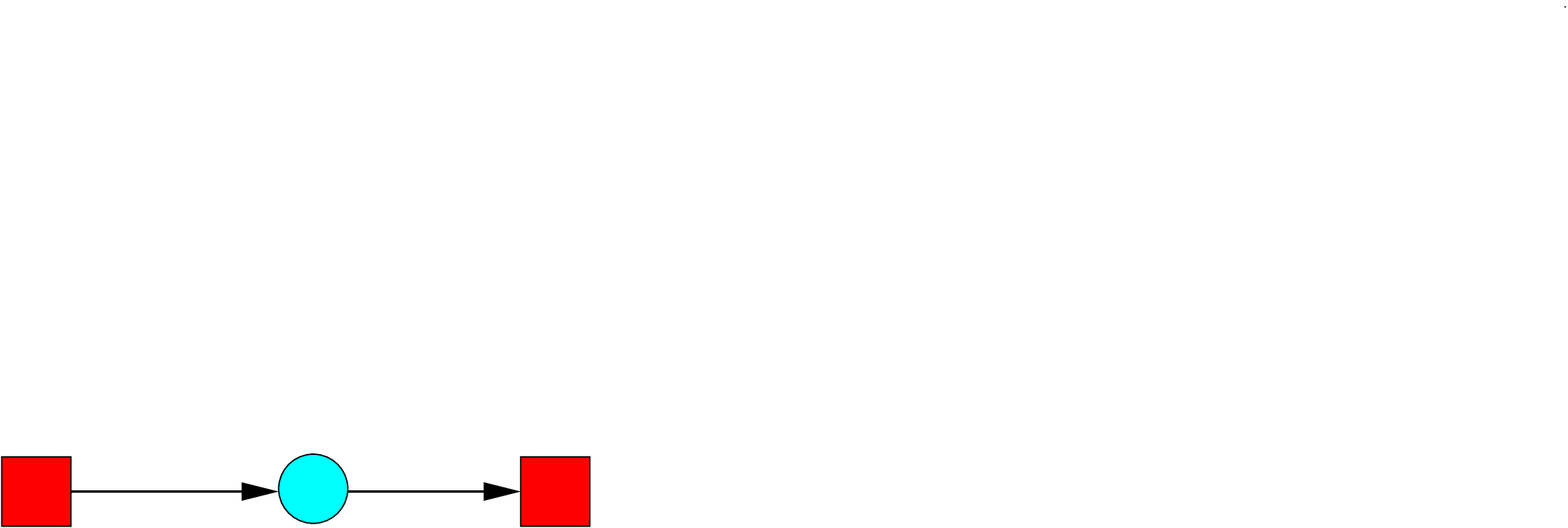}
}
\hspace{0.3in}
\subfigure[] 
{
    \label{case3}

\includegraphics[width=1in]{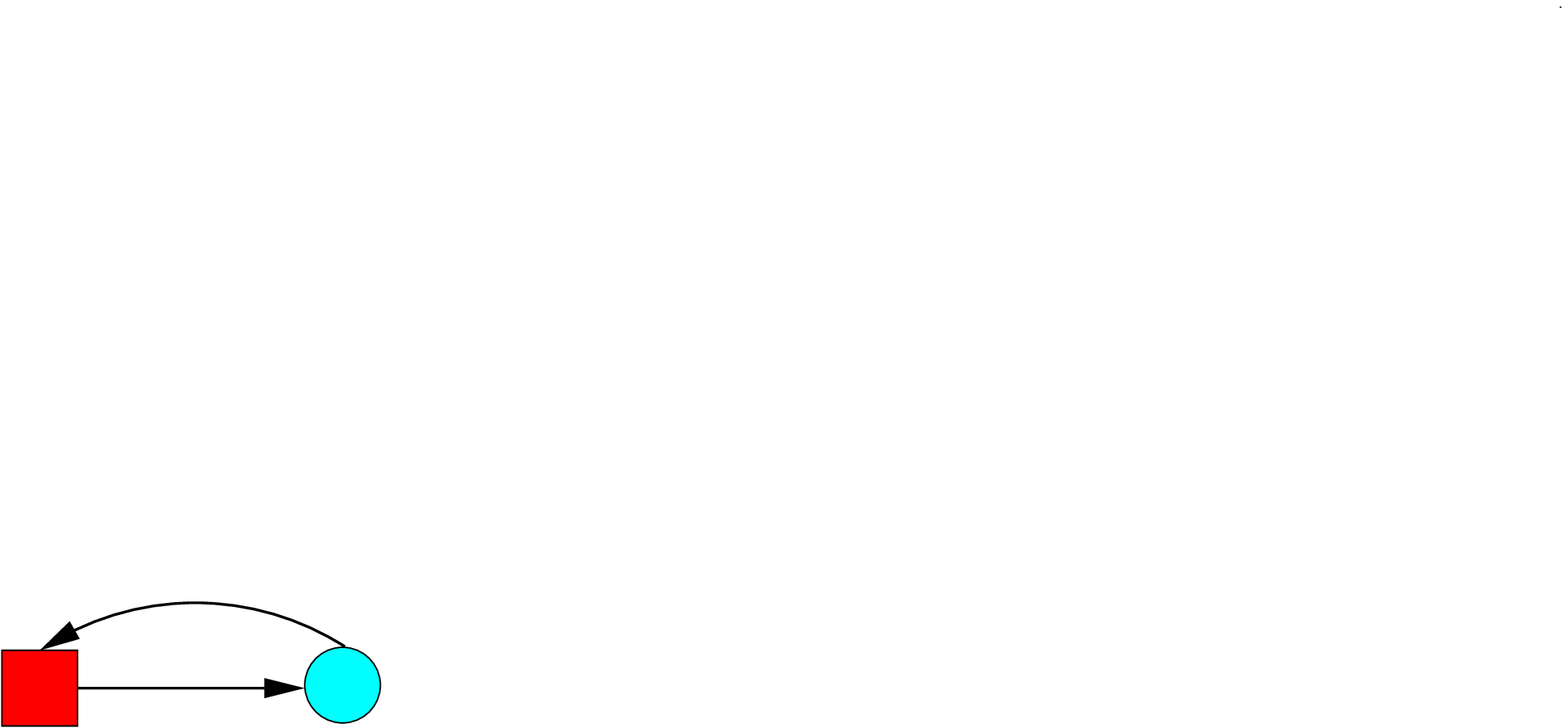}
}
\hspace{0.3in}
\subfigure[] 
{
    \label{case23}

\includegraphics[width=1.4in]{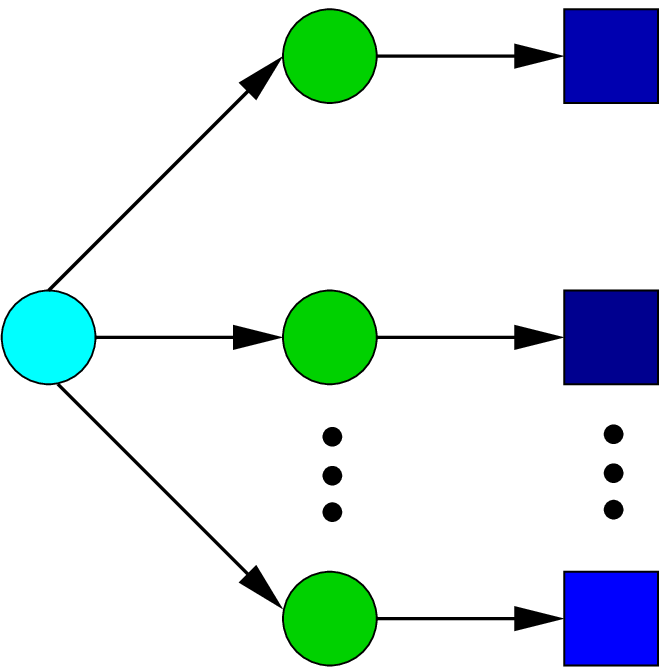}
}
\hspace{1.43in}
\subfigure[] 
{
    \label{instanton}

\includegraphics[width=2.2in]{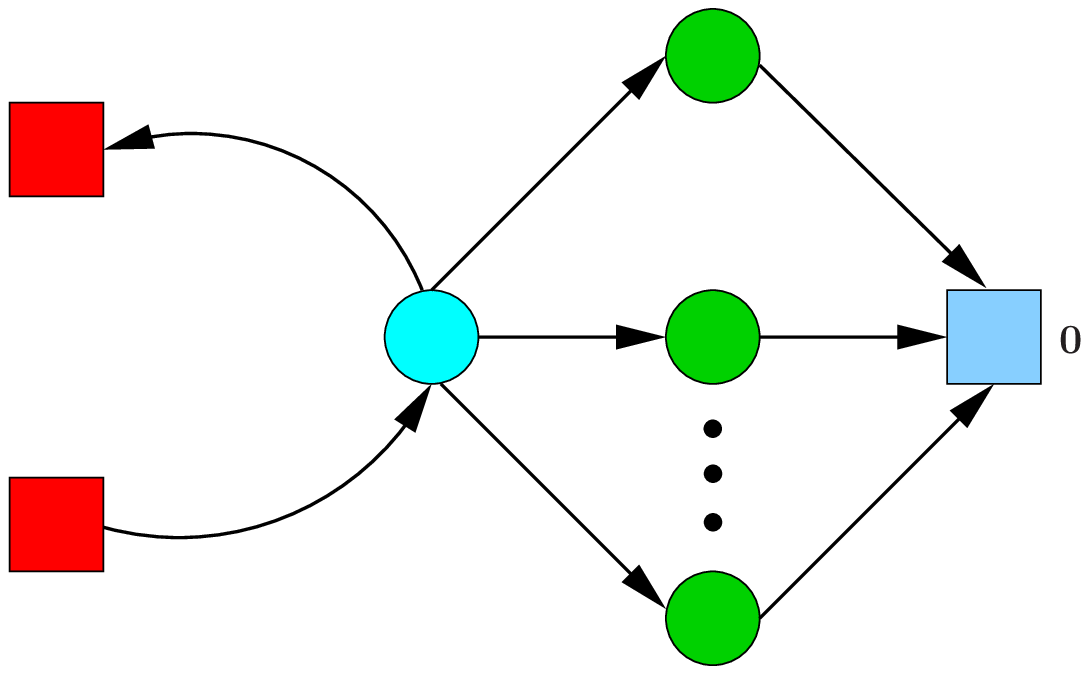}
} \caption{Squares represent pseudo-codewords and circles represent medians or
related noise configurations \subref{case1} LP decodes median of a
pseudo-codeword into another pseudo-codeword of smaller weight \subref{case2}
LP decodes median of a pseudo-codeword into another pseudo-codeword of the same
weight \subref{case3} LP decodes median of a pseudo-codeword into the same
pseudo-codeword \subref{case23} Reduced subset (three different green circles)
of a noise configuration (e.g. of a median from the previous step of the ISA)
is decoded by the LP decoder into three different pseudo-codewords
\subref{instanton} LP decodes the median (blue circle) of a pseudo-codeword
(low red square) into another pseudo-codeword of the same weigh (upper red
square). Reduced subset of the median (three configurations depicted as green
circles are all decoded by LP into all-zero-codeword. Thus,  the median is an
instanton.} \label{illustration}
\end{figure*}

\section{Instanton Search Algorithm and its Analysis}\label{algorithm}

In this Section, we describe the Instanton Search Algorithm. The
algorithm starts with a random binary vector with some number of flips
and outputs an instanton.

\noindent \underline{\textbf {Instanton Search Algorithm}}\\
\underline{\textit{Initialization (l=0) step}}: Initialize to a binary input vector $\mathbf{r}$ containing sufficient number of flips so that the LP decoder decodes it into a pseudo-codeword different from the  all-zero-codeword. Apply the LP decoder to $\mathbf{r}$ and denote the pseudo-codeword output of LP by $\mathbf{p}^{1}$.\\
\underline{\textit{$l\geq 1$ step}}: Take the pseudo-codeword $\mathbf{p}^l$
(output of the $(l-1)$ step) and calculate its median $M(\mathbf{p}^l)$. Apply
the LP decoder to $M(\mathbf{p}^l)$ and denote the output by
$\mathbf{p}_{M_l}$.  By Lemma \ref{lemma4}, only two cases arise:
\begin{itemize}
\item $w_{BSC}(\mathbf{p}_{M_l}) < w_{BSC}(\mathbf{p}^l)$. Then $\mathbf{p}^{l+1}=\mathbf{p}_{M_l}$ becomes the $l$-th step output/$(l+1)$ step input.

\item $w_{BSC}(\mathbf{p}_{M_l})=w_{BSC}(\mathbf{p}^l)$. Let the support of $M(\mathbf{p}^l)$ be $S=\{i_1,\ldots,i_{k_l}\}$. Let $S_{i_t}=S \backslash \{i_t\}$ for some $i_t \in S$. Let $\mathbf{r}_{i_t}$ be a binary vector with support $S_{i_t}$. Apply the LP decoder to all $\mathbf{r}_{i_t}$ and denote the $i_t$-output by $\mathbf{p}_{i_t}$. If $\mathbf{p}_{i_t}=\mathbf{0}, \forall i_t$, then $M(\mathbf{p}^l)$ is the desired instanton and the algorithm halts. Else, $\mathbf{p}_{i_t} \neq \mathbf{0}$ becomes the $l$-th step output/$(l+1)$ step input. (Notice, that Lemma \ref{lemma3} guarantees that any $\mathbf{p}_{i_t} \neq \mathbf{p}^{l}$, thus preventing the ISA from entering into an infinite loop.)
\end{itemize}

\noindent

Fig. \ref{illustration} illustrates different scenarios arising  in the
execution of the ISA. Here, the squares represent pseudo-codewords and the
circles represent binary vectors (noise configurations). Two squares of the
same color have identical pseudo-codeword weight and two circles of the same
color consist of same number of flips. Fig. \ref{case1} shows the case where a
median, $M(\mathbf{p}^l)$, of a pseudo-codeword $\mathbf{p}^l$ converges to a
pseudo-codeword $\mathbf{p}_{M_l}$ of a smaller weight. In this case,
$\mathbf{p}^{l+1}=\mathbf{p}_{M_l}$. Fig. \ref{case2} illustrates the case
where a median, $M(\mathbf{p}^l)$, of a pseudo-codeword $\mathbf{p}^l$
converges to a pseudo-codeword $\mathbf{p}_{M_l}$ of the same weight. Fig.
\ref{case3} illustrates the case where a median, $M(\mathbf{p}^l)$, of a
pseudo-codeword $\mathbf{p}^l$ converges to the pseudo-codeword
$\mathbf{p}^{l}$ itself. In the two latter cases, we consider all the binary
vectors whose support sets are subsets of the support set of $M(\mathbf{p}^l)$
and the vectors contain one flip less. We run the LP decoder with the vectors
as inputs and find their corresponding pseudo-codewords. One of the non-zero
pseudo-codewords found is chosen at random as $\mathbf{p}^{l+1}$. This is illustrated in
Fig. \ref{case23}. Fig. \ref{instanton} shows the case when all the subsets of
$M(\mathbf{p}^l)$ (reduced by one flip) converge to the all-zero-codeword.
$M(\mathbf{p}^l)$ itself could converge to $\mathbf{p}^l$ or some other
pseudo-codeword of the same weight. In this case, $M(\mathbf{p}^l)$ is an
instanton constituting the output of the algorithm.

We now prove that the ISA terminates (i.e., outputs an instanton) in the number of steps of the order the number of
flips in the initial noise configuration.
\begin{te}\label{theorem1}
$w_{BSC}(\mathbf{p}^{l})$ and $|\supp(M(\mathbf{p}^l))|$ are monotonically
decreasing. Also, the ISA terminates in at most $2k_0$ steps, where $k_0$ is the number of flips in the input.
\end{te}
\begin{proof}
If $\mathbf{p}^{l+1}=\mathbf{p}_{M_l}$, then $w_{BSC}(\mathbf{p}^{l+1}) <
w_{BSC}(\mathbf{p}^{l})$. Consequently, $|\supp(M(\mathbf{p}^{l+1}))| \leq
|\supp(M(\mathbf{p}^l))|$.

If $\mathbf{p}^{l+1}=\mathbf{p}_{i_t}$, then $w_{BSC}(\mathbf{p}_{i_t}) \leq 2
(|\supp(M(\mathbf{p}^l))|-1) < w_{BSC}(\mathbf{p}^l)$. Consequently,
$|\supp(M(\mathbf{p}^{l+1}))| \leq |\supp(M(\mathbf{p}^l))|$.

Since $w_{BSC}(\mathbf{p}^{j})$ is strictly decreasing, the weight of
pseudo-codeword at step $l$ decreases by at least one compared to the weight of
the pseudo-codeword at step $l-1$. Since by Lemma \ref{lemma1},
$w_{BSC}(\mathbf{p}^{1})\leq 2k_0$, the algorithm can run for at most
$2k_0$ steps.
\end{proof}
\textit{Remarks:} (1) By ``sufficient number of flips'', we mean that the initial binary vector should be noisy enough to converge to a pseudo-codeword other than the all-zero-codeword. While any binary vector with a large number of flips is almost guaranteed to converge to a pseudo-codeword different from the all-zero-codeword, such a choice might also lead to a longer running time of the ISA (from Theorem \ref{theorem1}). On the other hand, choosing a binary vector with a few number of flips might lead to convergence to the all-zero-codeword very often, thereby necessitating the need to run the ISA for a large number of times. 

(2) Theorem \ref{theorem1} does not claim that the algorithm finds the minimum
weight pseudo-codeword or the instanton with the smallest number of flips.
However, it is sometimes possible to verify if the algorithm has found the
minimum weight pseudo-codeword. Let $w_{min}^{ISA}$ denote the weight of the
minimum weight pseudo-codeword found by the ISA. If $w_{min}^{ISA}= 2\lceil
d_{frac}/2 \rceil -1$, then $w_{min}^{ISA}=w_{min}^{BSC}$.

(3) At some step $l$, it is possible to have $w_{BSC}(\mathbf{p}_{M_l}) =
w_{BSC}(\mathbf{p}^l)$ and incorporating such pseudo-codewords into the
algorithm could lead to lower weight pseudo-codewords in the next few steps.
However, this inessential modification was not included in the ISA to
streamline the analysis of the algorithm.

(4) While we have shown that $w_{BSC}(\mathbf{p}^l)$ decreases by at least
unity at every step, we have observed that in most cases, it decreases by at
least two. This is due to the fact that the pseudo-codewords with odd weights
outnumber pseudo-codewords with even weights. As a result, in most cases, the
algorithm converges in less than $k_0$ steps. (For illustration of this point
see example discussed in the next Section.)

(5) At any step, there can be more than one median, and the ISA does not specify which one to pick. Our current implementation
suggests to pick a median at random. Also, the algorithm does not provide
clarification on the choice of the pseudo-codeword for the case when more than
one noise configurations from the subset $\mathbf{r}_{i_t}$ converge to
pseudo-codewords distinct from the all-zero-codeword.  In this degenerate case,
we again choose a pseudo-codeword for the next iteration at random. Note that
one natural deterministic generalization of the randomized algorithm consists
of exploring all the possibilities at once. In such a scenario, a tree
of solutions can be built, where the root is associated with one set of initiation flips, any
branch of the tree relates to a given set of randomized choices (of medians and
pseudo-codewords), and any leaf corresponds to an instanton.

\section{Numerical Results}\label{results}
\begin{figure*}
\centering
\includegraphics{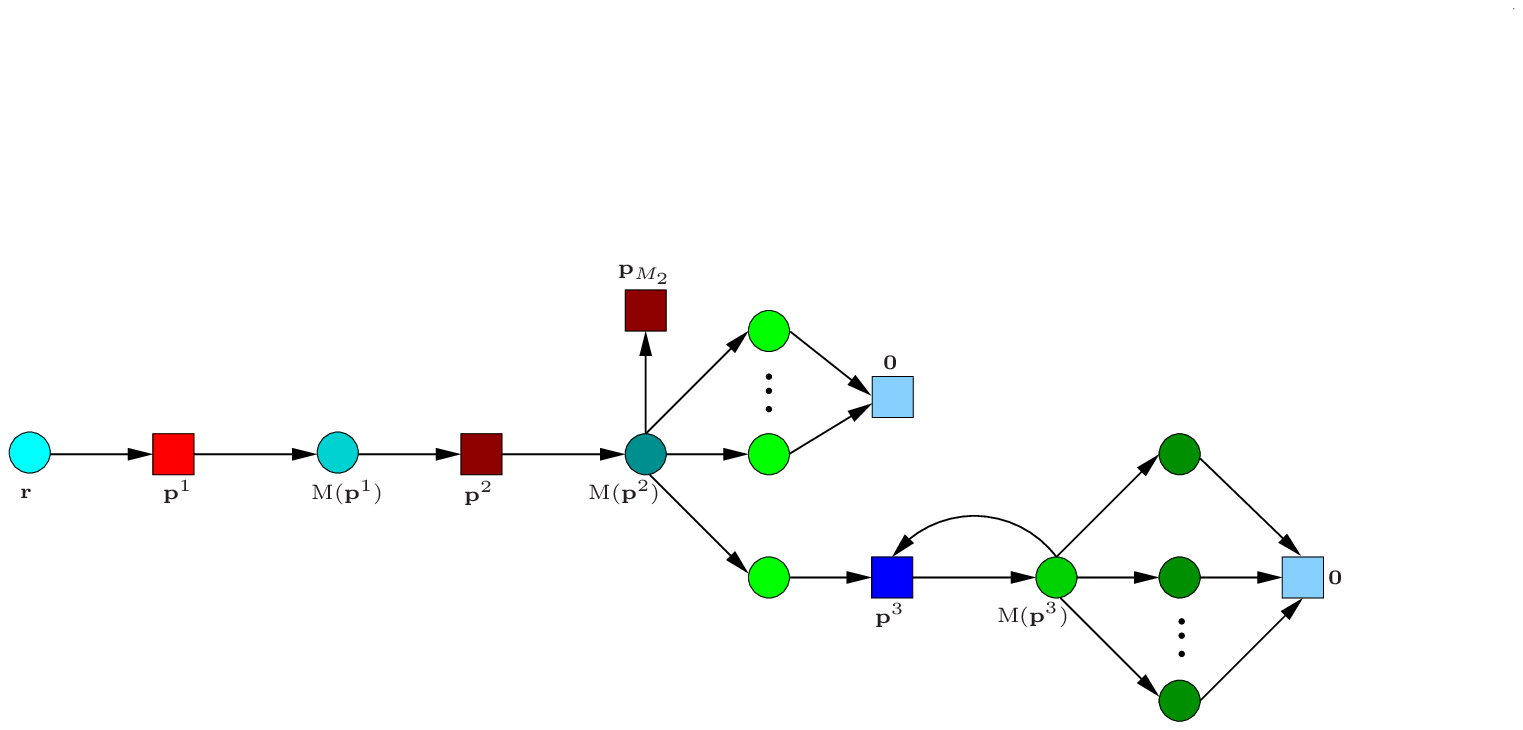}
\caption{Illustration for the example of ISA execution on the $[155,65,20]$ Tanner code discussed in Section \ref{results}.}
\label{completeexample}
\end{figure*}

In this Section, we present results illustrating different aspects and features
of the ISA. We use the $[155,64,20]$ Tanner code \cite{tannercode} for 
illustration purposes. We begin with an actual (and rather typical) example.
The reader is advised to follow this example with an eye on Fig.
\ref{completeexample}.

\begin{ex} \label{ex1}
The algorithm is initiated with a binary vector $\mathbf{r}$ whose support set
has cardinality $12$.  In this case, $\mathbf{r}$
converges to a pseudo-codeword $\mathbf{p}^1$ of weight 17 (Lemma \ref{lemma1} guarantees that $w_{BSC}(\mathbf{p}^1)\leq 24$). The Median
$M(\mathbf{p}^1)$ of the pseudo-codeword $\mathbf{p}^1$  has $9$ flips.
$M(\mathbf{p}^1)$ converges to a pseudo-codeword $\mathbf{p}_{M_1}$ of weight
$11$, marked as $\mathbf{p}^2$, whose median $M(\mathbf{p}^2)$ contains $6$
flips. $M(\mathbf{p}^2)$ decodes to a pseudo-codeword $\mathbf{p}_{M_2}$ of
weight $11$ and hence we consider all vectors whose support sets consist of one
flip less than in the support set of $M(\mathbf{p}^2)$. There are $6$ such
vectors and $5$ of them decode to the all-zero-codeword (we do not show all the
six vectors in Fig .\ref{completeexample}). The remaining vector decodes to a
pseudo-codeword of weight $9$, marked as $\mathbf{p}^3$. The pseudo-codeword
$\mathbf{p}^3$ has only one median $M(\mathbf{p}^3)$ which is decoded to the
same pseudo-codeword $\mathbf{p}^3$. Hence, we consider all (five) vectors
built from  the median $M(\mathbf{p}^3)$ removing a single flip and observe
that the LP decoder decodes all these vectors into the all-zero-codeword. We
conclude that the median is actually an instanton of size $5$.
\end{ex}

\begin{figure*}
\centering
\vspace{0.1in}
\includegraphics[width=0.9\textwidth]{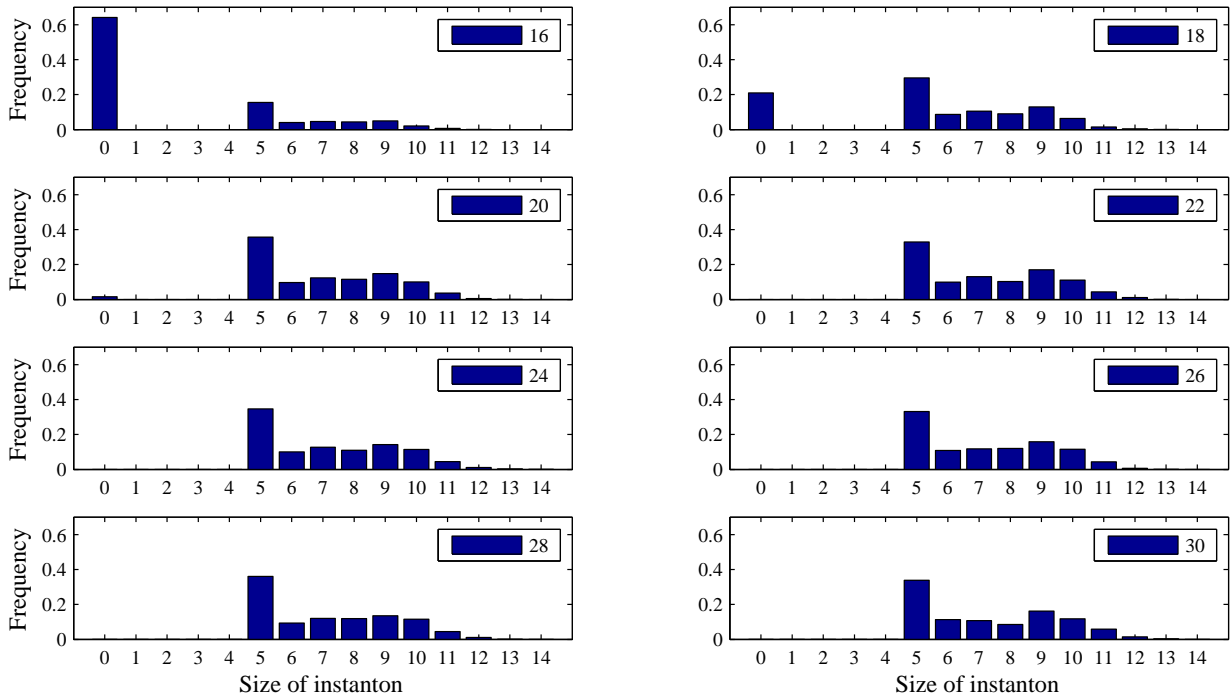}
\caption{Frequency of instanton (ISA outputs) sizes for different weights of
inputs. (Total length of the bars (for any of the sub-plots) should sum to
one.) The bar-graphs were obtained by running the ISA for 2000 different random
initiations with the fixed number of flips (ranging from $16$ to $30$). Numbers
at zero (if any) show the frequency of patterns decoded into the
all-zero-codeword.} \label{frequency}
\end{figure*}

We ran $2000$ ISA trials using random inputs with fixed number of initiation
flips. Fig. \ref{frequency} shows the frequency of the instanton sizes for the
number of initiation flips ranging from $16$ to $30$. The value at zero should
be interpreted as the number of patterns that decode to the all-zero-codeword.
It can be seen that if the initial noise vector consists of $22$ or more flips,
then it converges to a pseudo-codeword different from the all-zero-codeword in
all of $2000$ cases.

\begin{figure*}
\centering
\vspace{0.1in}
\includegraphics[width=0.9\textwidth]{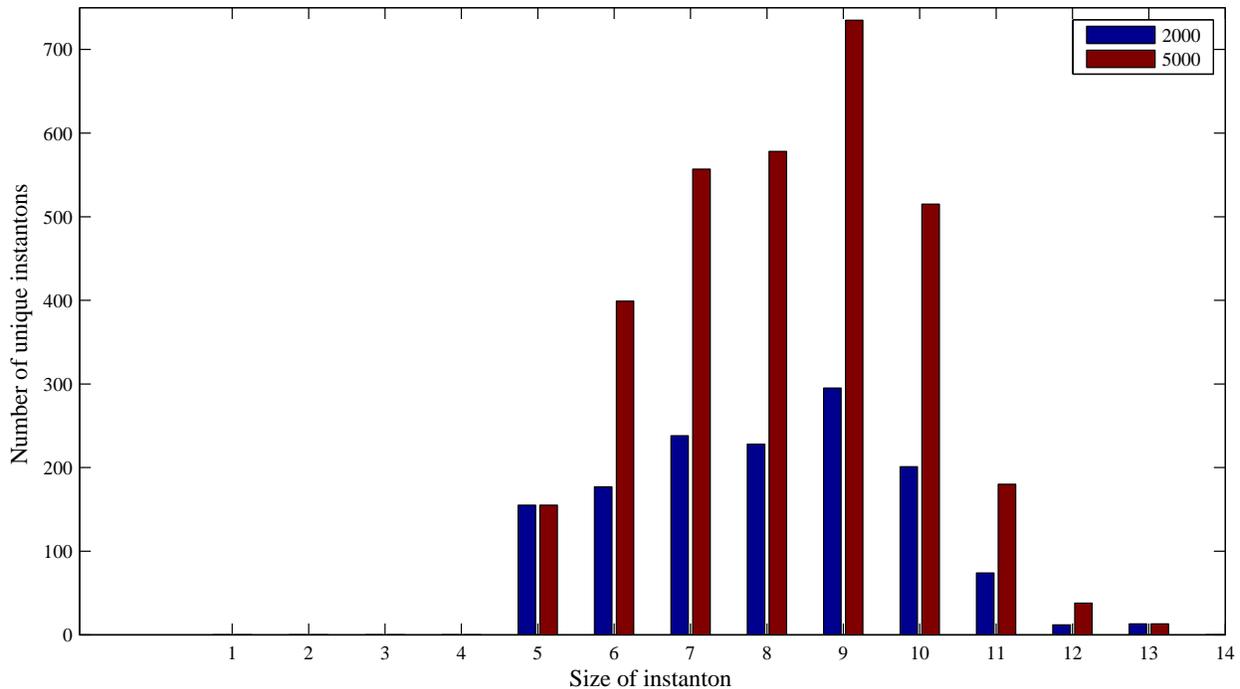}
\caption{Instanton-Bar-Graph showing the number of unique instantons of a given
weight found by running the ISA with 20 random flips for 2000 and 5000
initiations respectively.} \label{IBA}
\end{figure*}

Note that Fig.~\ref{frequency} shows a count of the total number of instantons of the same size, so
that multiple trials of ISA may correspond to the same instanton. To correct
for this multiplicity in counting, one can also find it useful (see discussions
below) to study the total number of unique instantons observed in the ISA
trials, coined the Instanton-Bar-Graph. Fig.~\ref{IBA} shows the number of
distinct instantons of a given size for $2000$ and $5000$ random initiations
with $20$ flips. One finds that the total number of ISA outputs of size $5$
after $2000$ trails is $720$, however representing only $155$ distinct
instantons. In this case (of the Tanner code), we can independently verify
\footnote{We have observed that all the instantons of size $5$ are in fact the
$(5,3)$ trapping sets described in \cite{06CSV}. Further investigation of the
topological structure of instantons will be dealt in future work.} that the
total number of  instantons of size $5$ is indeed $155$, thus confirming that
our algorithm has found all the instantons of length $5$ detecting each of them
roughly $4$ times. Obviously, the total number of distinct instantons of size
$5$ does not change with further increase in the number of trails. This
observation emphasizes utility of the sub-plots, with different number of
initiations, as the comparison allows to judge the sufficiency (or insufficiency)
of the number of trials for finding all the given size instantons. Extending
the comparison to larger size ($>5$) instantons, one observes that the numbers
change in transition from $2000$ to $5000$ trials,  thus indicating that
the statistics  is insufficient (at least after $2000$ trials) as some of the
instantons have not been found yet.

The smallest weight instanton found by the ISA is $5$. The accuracy of this
estimate can be verified (indirectly) by finding the $d_{frac}$ of the code.
Using the method outlined in \cite{05FWK}, we observed that $d_{frac}$ of the
Tanner code is $8.3498$. This implies that $w_{BSC}^{min} \geq 9$ (by Lemma
\ref{lemma0}), which in turn implies that the size of any instanton cannot be
less than $5$. This proves that here $5$ is, indeed, the smallest instanton
size, and respective minimum pseudo-codeword weight is $9$. Note also that the
fractional weight of all the $155$ pseudo-codewords of weight $5$ is $9.95$,
while the weight of the pseudo-codeword with the minimal fractional weight of
$8.3498$ is $19$. The remark illustrates that minimality of the fractional
weight does not imply minimality of the pseudo-codeword weight (and thus
minimality of the respective instanton size).

\section{Summary and Open Problems}\label{conclusion}

In this paper, we characterized failures of the LP decoder over the BSC in
terms of the instantons and respective pseudo-codewords. We then provided an
efficient algorithm for finding the instantons.  The ISA is guaranteed to terminate in the number of steps upper bounded by
twice the number of flips in the original input (Theorem 1). Repeated
sufficient number of times, the ISA outcomes the Instanton-Bar-Graph showing
the number of unique instantons of different sizes. We also proved that the LP
decoding of any configuration of the input noise which includes an instanton
leads to a failure (Lemma 7). This Lemma arguably suggests to use the
Instanton-Bar-Graph derived with the ISA algorithm as a metric for code
optimization.

Finally, we conclude with an incomplete list of open problems and directions
for future research following from this study:

(1) One would like to understand how to choose initiation of the ISA which
guarantees convergence to the smallest size instanton.

(2) When can one be reasonably certain that all instantons of a given weight
are found? Or stating it differently,  how many trials of the ISA are required
to find all the instantons of the given size? Does the number of trials scales
linearly with the size of the code?

(3) We have noticed that difficulty of finding an instanton grows with its
size.  Once the ISA finds all the instantons of certain weight, can one
optimize initiation strategy for the algorithm to find instantons of larger
size more efficiently?

(4) Can one utilize knowledge of the code structure (e.g. for highly structured
codes) to streamline discovery of the Instanton-Bar-Graph,  especially in the
part related to the larger size instantons?

(5) Some studies have explored connections between pseudo-codewords and
stopping sets (see e.g. \cite{kellysridhara}). Are there any (similar?)
relationships between trapping sets of the BSC (for Gallager like algorithms)
and BSC-LP instantons?

(6) Are instantons of a code performing over the BSC related to instantons of
the same code over the AWGN channel (or other soft channels)? Can we use one to
deduce the other?
\section*{Acknowledgment}
The authors would like to thank P. Vontobel for his comments and suggestions and D. Sridhara for his clarifications regarding Lemma 1. The work by S. K Chilappagari was performed when he was a summer GRA at LANL. The work at LANL, by S. K. Chilappagari and M. Chertkov, was carried out under the auspices of the National Nuclear Security Administration of the U.S. Department of Energy at Los Alamos National Laboratory under Contract No. DE-AC52-06NA25396. B. Vasic  would like to acknowledge the financial support of the NSF and Seagate Technology.\bibliographystyle{IEEEtran}

\end{document}